\documentstyle[epsfig]{aipproc}
\def\bea{\begin{eqnarray}}
\def\eea{\end{eqnarray}}
\def\ra{\rightarrow}
\def\be{\begin{equation}}
\def\ee{\end{equation}}

\def\eg{{\it{e.g. }}}
\def\ie{{\it{i.e. }}}
\def\g2{GeV$^2$}

\def\NPB{{\em Nucl. Phys.} B}

\def\PLB{{\em Phys. Lett.}  B}
\def\PRL{{\em Phys. Rev. Lett.} }
\def\PRD{{\em Phys. Rev.} D}
\def\ZPC{{\em Z. Phys.} C}
\def\PRC{{\em Phys. Rep.} C}

\def\EPJC{{\em Eur. Phys. J. }C}

\def\JPG{{\em J. Phys.} G}  
\def\PR{{\em Phys. Rep. }}

\begin{document}
\begin{flushright}
{\sl All the fifty years of conscious brooding\\
 have brought me no closer to the answer to the question,\\ 

{ ``What are light quanta?''} \\

Of course today every rascal thinks he knows the answer,\\
but he is deluding himself.\\
{\em A. Einstein, 1951}}
\end{flushright}
\title{Structure Functions for the Virtual and   Real Photons\thanks{
 Invited talk at the International Conference on 
The Structure and Interactions of the Photon
Including the 13th International Workshop on 
Photon-Photon Collisions, PHOTON 2000, Ambleside, England, 26-31 August 2000}}
\author{Maria Krawczyk \thanks{Partly supported by KBN Grant No 2P03B0511(2000)
}}
\address{Institute of Theoretical Physics, Warsaw University,
Warsaw, Poland}
\maketitle
\begin{abstract}
Development  of concepts related to the photon and  its
high energy hadronic interaction is briefly reviewed.
A photon considered as an ideal probe of hadron structure,
paradoxically   is also considered as an ideal target
to test the perturbative QCD.
The present status of the theoretical and  experimental analysis
of the structure functions for a virtual and real photon  is presented. 
\end{abstract}
\section*{100 years of light quanta}
The photon is the best known boson and one of the oldest elementary 
particles.  This year is special - we celebrate  100 years of light quanta;
 below I list the  crucial  dates in 
the development of notion of the photon \cite{100}:\\
\noindent
$\bullet$ 1900 - Planck's hypothesis of quantum of electromagnetic energy,
 $E=h \nu$\\
$\bullet$ 1905 - Einstein's hypothesis of  quantum of light ($\gamma$),
 $E=h\nu=pc$ \\                           
$\bullet$ 1915 - Millikan's  experiment: photo-emission from metal\\
$\bullet$ 1922 - Compton  experiment: $\gamma e \ra \gamma e$    \\
In the next years (1925-7)  the Quantum Electrodynamics
(QED) was invented  by Born, Heisenberg, Jordan, Dirac and others 
\cite{schweber},   
with photon  playing the  role 
of  a gauge particle of electromagnetic interaction.  
Few years later (1931) Wigner gave the complete group theoretical description
of  angular-momentum states, according to which 
  photon means  helicity states of spin 1 massless particle\cite{kim}.  
The current name: the photon was given in 1926 by the  chemist G.N. Lewis
\cite{name}.
\subsection*{Photon-hadron interaction}
The photon  properties, as follows from Quantum Electrodynamics, 
are well known: the photon is a  massless, chargeless object with a 
pointlike coupling to the charged  fundamental particles. As such, it  
is an ideal tool for  probing structure of more complicated objects,  
for example hadrons, acting as  a microscope with the resolution given by its 
wavelength.

However, in the high energy photon-hadron interaction there are phenomena 
which can (should ?) be interpreted in terms of 
`` hadronic (partonic) structure'' of the {\sl photon}.
This way one can  effectively describe    leading contributions 
to the rates of certain processes. 
Some early ideas and facts   are recalled below:\\
\noindent
$\bullet$  1960-72 - observation of hadronic  properties of the photon in soft processes,
the $\rho (\omega,\phi)$-photon analogy,
 Vector Dominance Model (VDM) (also GVDM) \cite{vdm}\\
$\bullet$ 1969-71 - an importance of  $\gamma \gamma\ra hadrons$ processes 
in $e^+e^-$ collisions \cite{bud}\\
$\bullet$ 1970-74 - a  deep inelastic scattering on the real
photon \cite{walsh}a) and
Parton Model predictions
for structure functions $F_{1,2(L)}^{\gamma}$, and $F_{3}^{\gamma}$,  
 $g_1^{\gamma}$
  \cite{walsh}b) \\
$\bullet$ 1977-80 - 
 asymptotic (point-like) solution in QCD (LO \cite{witten} 
and NLO \cite{bb} results); structure functions of a real photon
 as a unique test of QCD\\
$\bullet$ 1979-84  - singularities in the asymptotic solutions 
for a real photon at small $x$;  negative $F_2^{\gamma}$
in the NLO QCD analysis \cite{bb,uematsu,bar,rossi} \\
$\bullet$ 1981-84 -  hadronic contribution to $F_2^{\gamma}$
 as a cure of the problem at small $x$ \cite{bar,rossi}\\   
$\bullet$ 1981-84 -  structure functions  
of virtual photons  (singularity free LO and NLO predictions) - 
a unique test of QCD \cite{uematsu,rossi} \\
$\bullet$ 1989-93 - relation of spin-dependent  structure functions of photons
 to QED and QCD  anomaly
\cite{efremov}.

Starting from 1981    structure functions of unpolarized real 
and virtual photons, $F_2^{\gamma}$ and $F_{eff}^{\gamma^*}$, respectively,
 are being measured.  Ten years later the first data on the 
production of large $p_T$ particles and jets  in the photon-induced processes, 
so called   resolved photon processes,   have appeared for both
 real and virtual photons. Review of data can be found in  
\cite{nisius},\cite{rev}.

\subsection*{The photon A.D.2000}

The above short outline of historical development of basic concepts  
related to the hadronic "structure" of the photon shows how in the past high 
expectations were followed by  deep defeats. Even nowadays the situation is far 
from being clear.
Although for many hadronic processes involving photons
there exist  already NLO QCD calculations, a 
 proper way of describing   the photon interaction is still 
a subject of ongoing discussions, \eg how to count  the order of the 
perturbation \cite{order}.
Still,  after so many years of photon physics  there is  a lot of confusion, 
even  terminology seems to be inadequate and generates additional problems, 
\eg \cite{chyla0}.
The fact that a photon has a double face - being a probe and a target,
sometimes in the same process, obviously  does not help.

The main source of data   on  the {strong (hadronic,partonic) 
properties} of the real photon comes from 
DIS$_{e\gamma}$ experiments in the $e^+e^-$ collisions.
Recent results on the structure function $F^{\gamma}_2$,
 based on few years runs' at LEP1
 and  TRISTAN  at the CM energy $\sim$ 90 and 60 GeV, respectively,
 are now available. There are also new data taken at LEP 
at higher  energies. Altogether the
existing data  cover a wide range of the (average) $Q^2$ from 0.2 to 
$\sim 706$ GeV$^2$ \cite{taylor}.
The range of the (center of bin) $x_{Bj}$ variable extends 
from $\sim$ 0.001 \cite{clay} to 0.98.
Recently the  first dedicated  
measurement of $F^{\gamma}_{2,c}$ has been performed at LEP2, see  
\cite{nisius-c}. In addition there exist data of the leptonic structure 
functions, see \eg \cite{nisius}. 

The complementary data on the photon structure
 are coming from  measurements
 of the resolved real photon processes, 
\ie production of large $p_T$ jets (also individual hadrons, photons, 
heavy quarks),    in ${\gamma}{\gamma}$ collisions at
$e^+e^-$ machines and in  the photoproduction  at the $e^{\pm}p$ 
collider HERA ($\sqrt s$ $\sim$ 300 - 320  GeV), see \cite{wing}.

The determination  of the hadronic structure function $F_2^{\gamma}$
for a real photon in $e^+e^-$ collision 
relies on the  unfolding, therefore precision of these data
 depends crucially  on the accurate description of the hadronic final state by 
the Monte Carlo models.
The improvement in the unfolding in DIS -type experiments 
has been obtained recently, still the 
dependence on the chosen MC used in the analysis
cannot be avoided \cite{LEPW}, see also \cite{miller}a-b). 
In the modeling of the final state, 
one  includes \cite{fkp,sas} various initial ``states'' of the target photon:
the photon as a $q \bar q$ pair and  the photon as a $\rho$ meson,
with a relevant structure. The  virtual photon (a probe)
 can interact    by its constituents as well, see \eg  \cite{kap}. 
So,  in the DIS-type 
analyses also processes with a resolved $\gamma$ ($\gamma^*$) are being  involved.

During the last years  data on 
 the ``structure of the virtual photon'' have appeared, mostly 
from the resolved virtual photon processes
at the $ep$ collider HERA (with the virtuality of the photon 
from    0.1 to 85 GeV$^2$), see \cite{wing}. 
Fifteen  years after the first measurement of the DIS$_{e\gamma^*}$ events
by PLUTO collaboration {\cite{PLUTO}},
a new measurement has just been performed  at LEP (L3 Coll.\cite{L3}).
Extraction of 
 the effective  structure function or of effective parton 
densities in a virtual photon from the $e^+e^-$ and $ep$ data
 is a  difficult task, especially if  
 interference terms are large as  observed in the OPAL experiment 
for double-tag leptonic events \cite{OPAL-qed,nisius}.

Basic  QCD predictions for processes with  real and virtual photon 
are definitely in agreement with  the data. 
However the discrepancies between the data and predictions of Monte Carlo 
models  for various distributions  in the photon-induced processes 
are observed both in $ep$ and $e^+e^-$ collision.
Implementation   of the modified transverse momenta distribution of 
the partons in the $\gamma$ and  taking into account 
multiple parton interaction  in the MC programs \cite{miller}c)
  help to describe the data. 
Nevertheless, the existing  data, also these on heavy quark production 
\cite{hq},
 seem to give us a message that the 
partonic content of the photon is not properly described 
by existing parametrizations.

Taking all these facts  into account I think it is 
sensible to  start from scratch with a basic introduction
to the concept of photon structure functions. I  use  as a model the 
``leptonic structure'' of a real photon. Next  I discuss 
  the  $e^+e^-$ environment of the DIS-type  experiments
for photons, where our basic knowledge comes from. At this stage the leptonic 
structure of the virtual photon can be introduced. Then the same steps
will lead us to the concept of  hadronic structure functions of the photon.

\section*{The "structure" of the light quanta}
In the quantum field theory, 
a photon, as any  elementary particle, can fluctuate into 
various states consisting of leptons, quarks, $W^{\pm}$ bosons, hadrons.
 {\sl  ``...through an interaction with a 
Coulomb field the photon could materialize as a pair of electrons,
$\gamma\rightarrow e^+e^-$. Although not usually thought of in these terms, 
this phenomenon was the earliest manifestation of photon structure''} 
\cite{vdm}b). 
\subsection*{Leptonic structure of the photon}
\subsubsection*{the target = $\gamma$}
One  can test leptonic structure  of photon  in  the
deep inelastic scattering, $e\gamma \ra e +~leptons$.
   Let us take a (unpolarized) real photon ($p^2=0$) as a target 
and assume that the probe, highly virtual photon $\gamma^*$, with large 
virtuality $-q^2=Q^2$, couples directly to the electric charge
of the fundamental particles. To describe this process 
one can introduce  the  corresponding structure functions, 
as for the proton case.  
The contribution of the lowest order  QED  process,  
 $ \gamma^*(q) \gamma (p) \ra l^+l^-$, Fig.1(left), to  $F_2^{\gamma}|_l$
is given by  (the {\it box} contribution) \cite{bud}: 
\begin{figure}[htb] 
\centerline{\epsfig{file=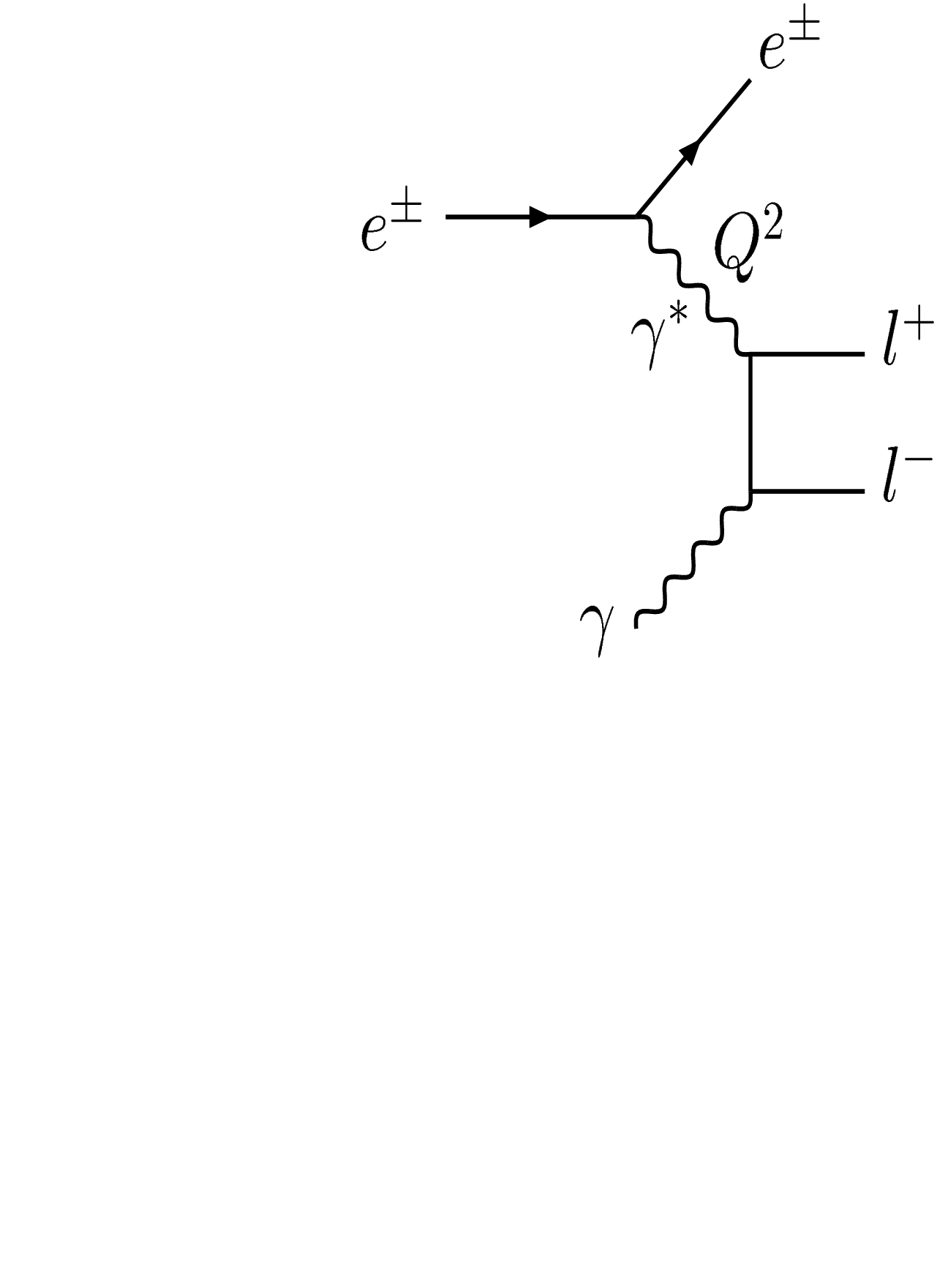,height=6cm,width=8cm}
\epsfig{file=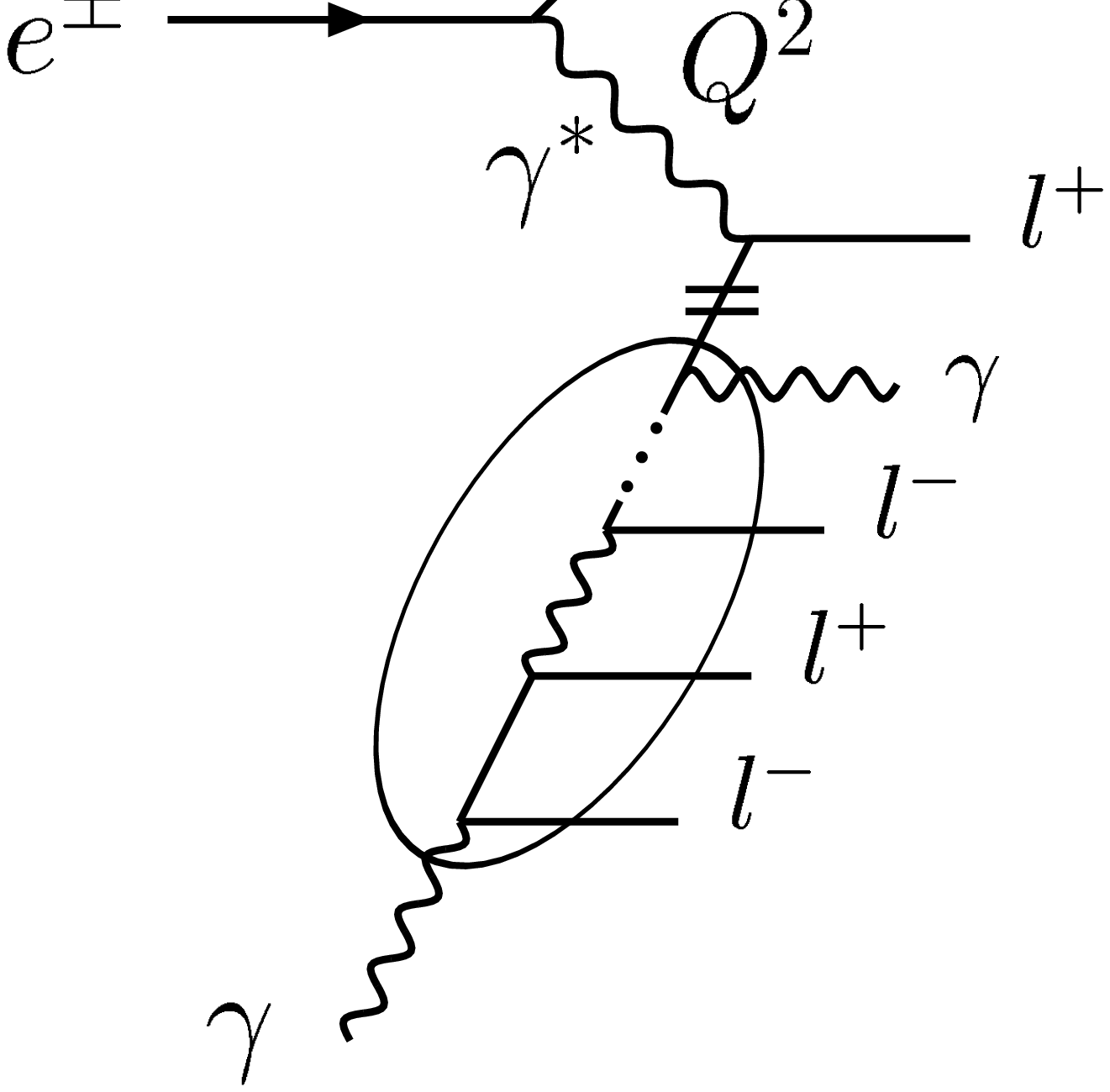,height=6cm,width=8cm}  
\epsfig{file=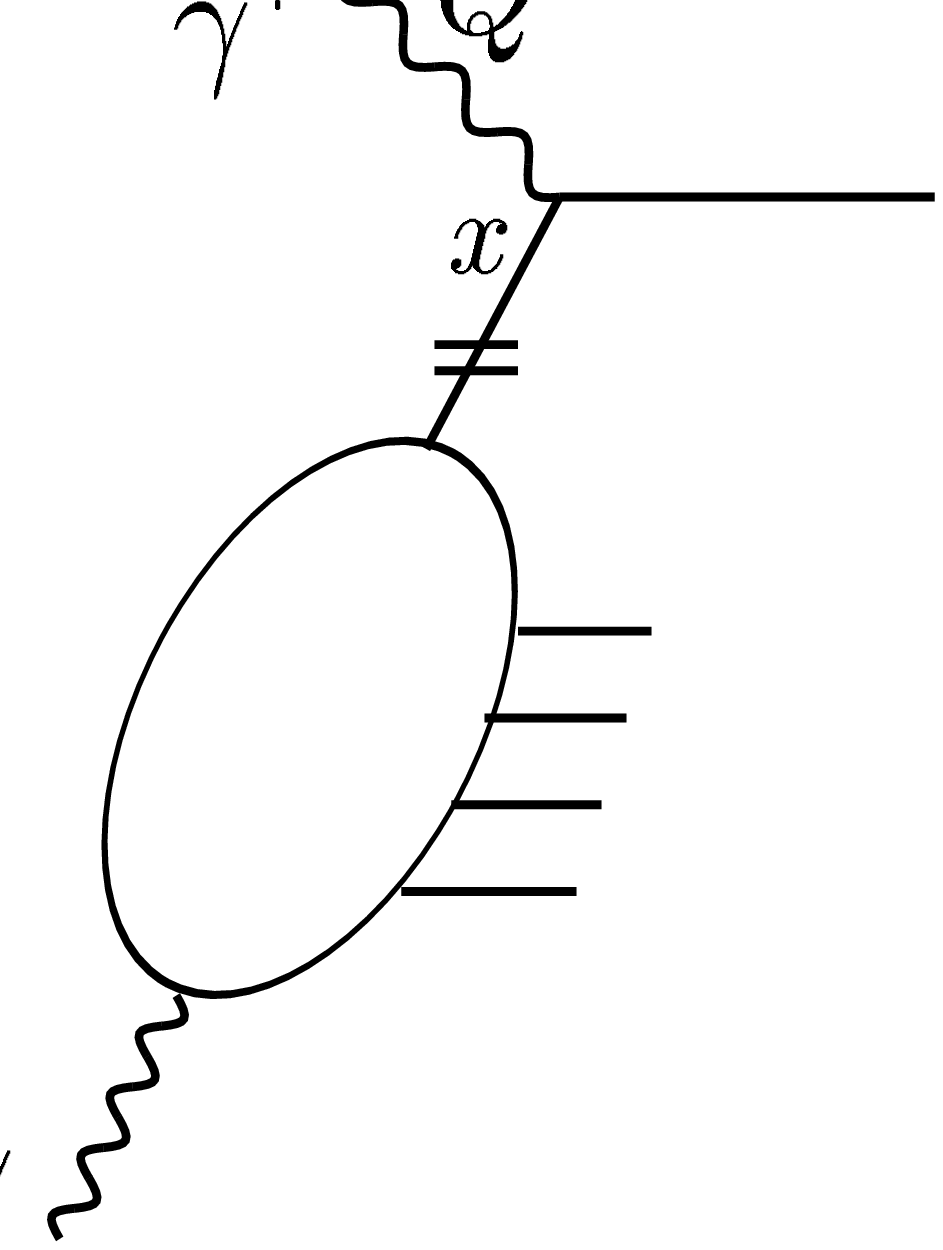,height=6cm,width=8cm}}
\vspace{-2.5cm}
\caption{The  lowest QED process for production of $l^+l^-$ 
final state (left);  the leptonic content of the photon in the LLA  (center);
a general representation of the DIS$_{e\gamma}$ with a point-like coupling
of the probe to the elementary constituent of the photon target with $x=x_{Bj}$ (right).} 
\end{figure} 
\begin{eqnarray}
\nonumber
{F}_2^{\gamma}|_{l}={{{\alpha}}\over{{\pi}}}
Q_l^4 x_{Bj}
[ (-1+8x_{Bj}(1-x_{Bj})-x_{Bj}(1-x_{Bj}){{4m_{l}^2}\over{Q^2}}){\beta}\\
{+[x_{Bj}^2+(1-x_{Bj})^2+x_{Bj}(1-3x_{Bj}){{4m_{l}^2}\over{Q^2}}
-x_{Bj}^2{{8m_{l}^2}\over{Q^2}}]\ln{{1+{\beta}}\over{1-{\beta}}} ] },
\eea
where  $x_{Bj}=Q^2/2pq$, $m_{l}$ -  lepton  mass, $Q_i$ - 
electric charge, $Q_l=1$, and $\beta$ is the lepton velocity in the $\gamma^* \gamma$ CM system.
\begin{figure}[b] 
\centerline{\epsfig{file=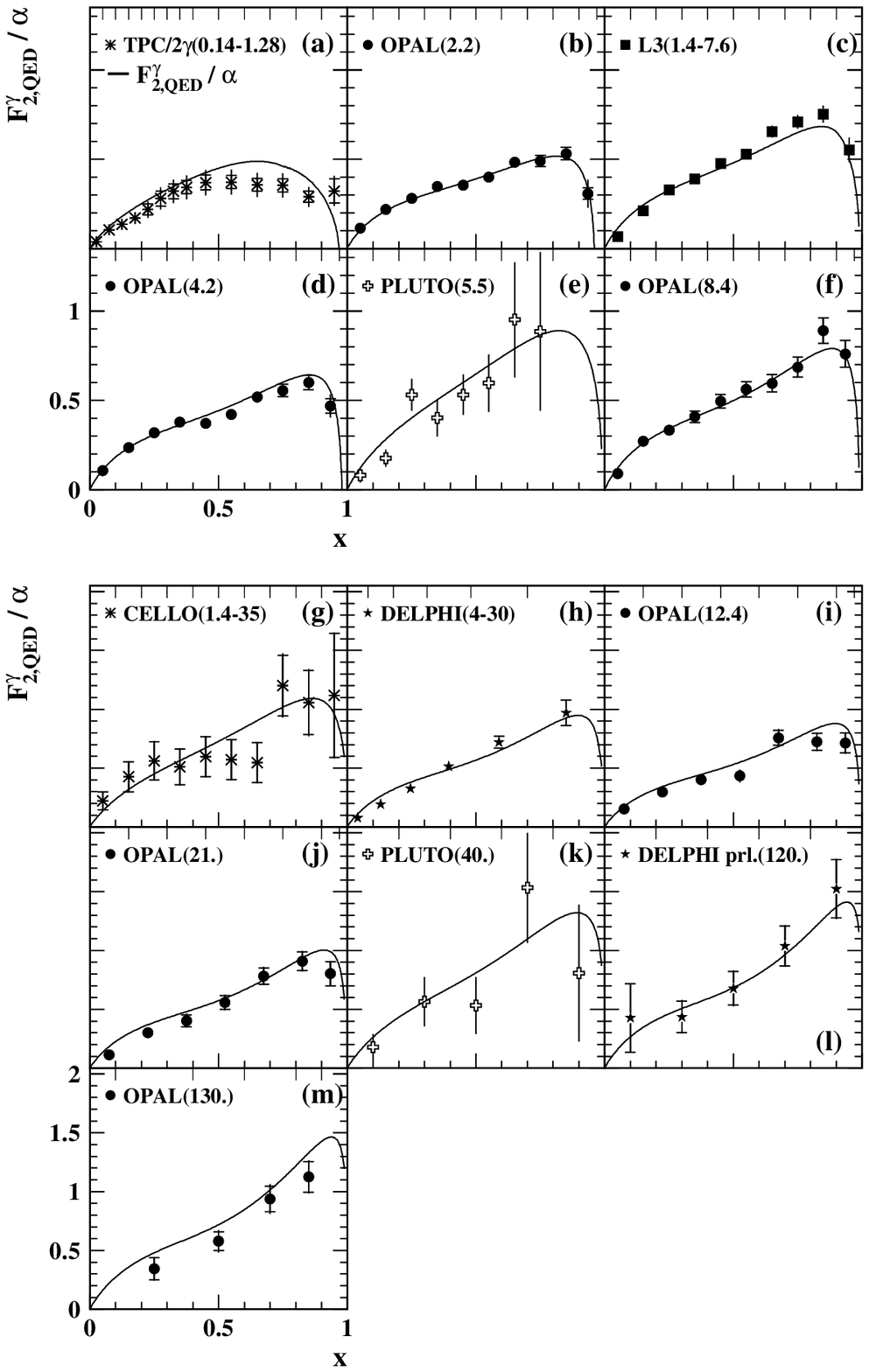,height=11cm,width=7.5cm}
\epsfig{file=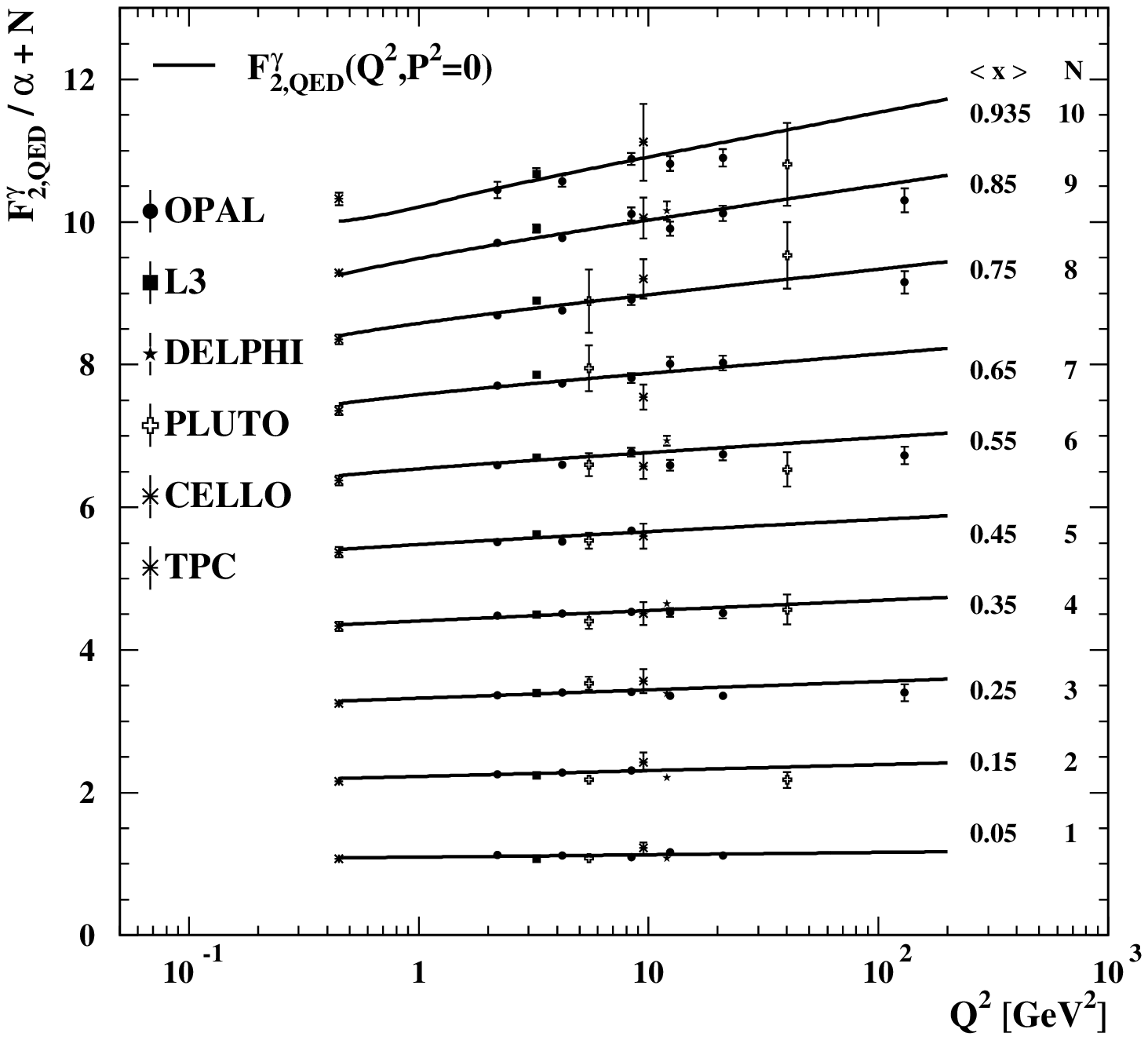,height=11cm,width=8cm}}  
\vspace{-1.5cm}
\caption{  Summary of existing $F_2^{\gamma}|_l/\alpha$ 
data, for $\mu^+\mu^-$ final state, shown with QED predictions for $P^2$=0,
 (left) - as a function of $x_{Bj}$
for broad $<Q^2>$ range, (right) - as a function of $Q^2$
for fixed $x_{Bj}$ bins, from [14]. }  
\end{figure}
For large  invariant mass of the $l^+l^-$ system,  $W\gg m_l$,
 $x_{Bj}$ not too close to 0 and 1, and neglecting terms 
$\sim  {{m_{l}^2}/{Q^2}}$ one gets (keeping the leading logarithmic (LL) term)
\be
{F}_2^{\gamma}|_{l}={{{\alpha}}\over{{\pi}}} 
Q_l^4 x_{Bj}
[x_{Bj}^2+(1-x_{Bj})^2]\ln{{Q^2}\over{m_{l}^2}}.
\ee
The obtained  $F_2^{\gamma}|_l$ is solely due to the QED interaction. 
It is large  at large $x_{Bj}$ and it  has 
the characteristic logarithmic rise with $Q^2$   from 
the {\it collinear} configuration in the $ \gamma \ra l^+ l^- $ splitting.  
In principle one can also introduce   a   {\it leptonic density}
in the real photon: in  the LLA  one gets $l^+(x_{Bj},Q^2)={{\alpha}\over{2\pi}}
Q_l^2 x_{Bj} [x_{Bj}^2+(1-x_{Bj})^2]\ln{{Q^2}\over{m_{l}^2}}$ (the same for 
$l^{-}$).
Here  $x_{Bj}=x$, where $x$ - the  part of the
four-momentum of the initial photon taken by  its  leptonic constituent
(Fig.1 (right)).

Leptonic structure of the  real photon  $F_2^{\gamma}|_l$ can be measured at 
future colliders, where  
beams  of energetic real photons can be obtained in the backward 
Compton scattering (Photon Colliders) \cite{telnov}.
Nowdays the   leptonic structure functions of a real 
photon are measured in various experiments at $e^+e^-$ colliders.
Here  (quasi) real photons with a Weizs\"acker-Williams energy 
spectrum play a role of a target.
The  $F_2^{\gamma}|_l$ data for  
 $l=\mu$ together with QED predictions   are summarized in Fig.2, from 
\cite{nisius}.

The basic features of the  lowest QED order result (2) will be modified 
only softly by higher QED corrections.
Formally leading logarithmic QED corrections,   
powers of $\alpha \log Q^2/m_e^2$, 
can be summed up using the {\it evolution equation} 
in $Q^2$ \cite{GL},
 with an  {\it inhomogenous term} due to the $\log Q^2$- dependence present
already in the  lowest order QED prediction (2).
  The resulting  {\it collinear QED cascade} included
in the LLA in the leptonic density $l^+(x,Q^2)$  
is  represented in Fig. 1(center). 
Here, starting from the first splitting of the initial photon
 all emission processes up to the interaction with a probe
are based  on point-like couplings.
\subsubsection*{the target = $\gamma^*$}
Leptonic structure functions can also be introduced for 
a virtual photon, \ie with $|p^2|=P^2 \neq 0$, to be 
measured in the lepton beams collision.
Let us discuss the production of an {\it arbitrary}   state $X$ 
in the process $e(p_1)e(p_2)\ra e(p_1')e(p_2')X$,
via the $\gamma^*(q_1) \gamma^*(q_2)$ collision
 ($ q_1=p_1-p_1'$,  $ q_2=p_2-p_2'$),
Fig.3 (left).
The corresponding cross section for the
unpolarized lepton beams,  assuming $\mid q_{1,2}^2 \mid \gg m_e^2$, 
and  typical conditions in present experiments, is given by 
(see \cite{bud}, also \cite{nisius})
\bea
E_1'E_2'{{d\sigma(ee\ra ee X)}\over{d^3p_1'd^3p_2'}} =
L_{TT}(\sigma_{eff}
+{1\over 2}\tau_{TT}\cos 2\bar{\phi}-
4 \tau_{TL}\cos \bar{\phi}),
\eea
where  helicity states of photons are denoted by  T - transverse 
(+ or -) and L - longitudinal (0). An effective
cross section is defined as  $\sigma_{eff}=\sigma_{TT}+\sigma_{LT}+\sigma_{TL}+\sigma_{LT}$, where
 $\sigma_{TT,TL,LT,LL}$ (the first subscript is  for the 
photon with $q_1$) denote 
the corresponding cross sections,  $\tau_{TT,TL}$ - 
the  interference terms.
The $\bar \phi$ is the angle between two scattering planes 
of the scattered electrons in the $\gamma^* \gamma^*$ CM system.
\begin{figure}[htb] 
\centerline{\epsfig{file=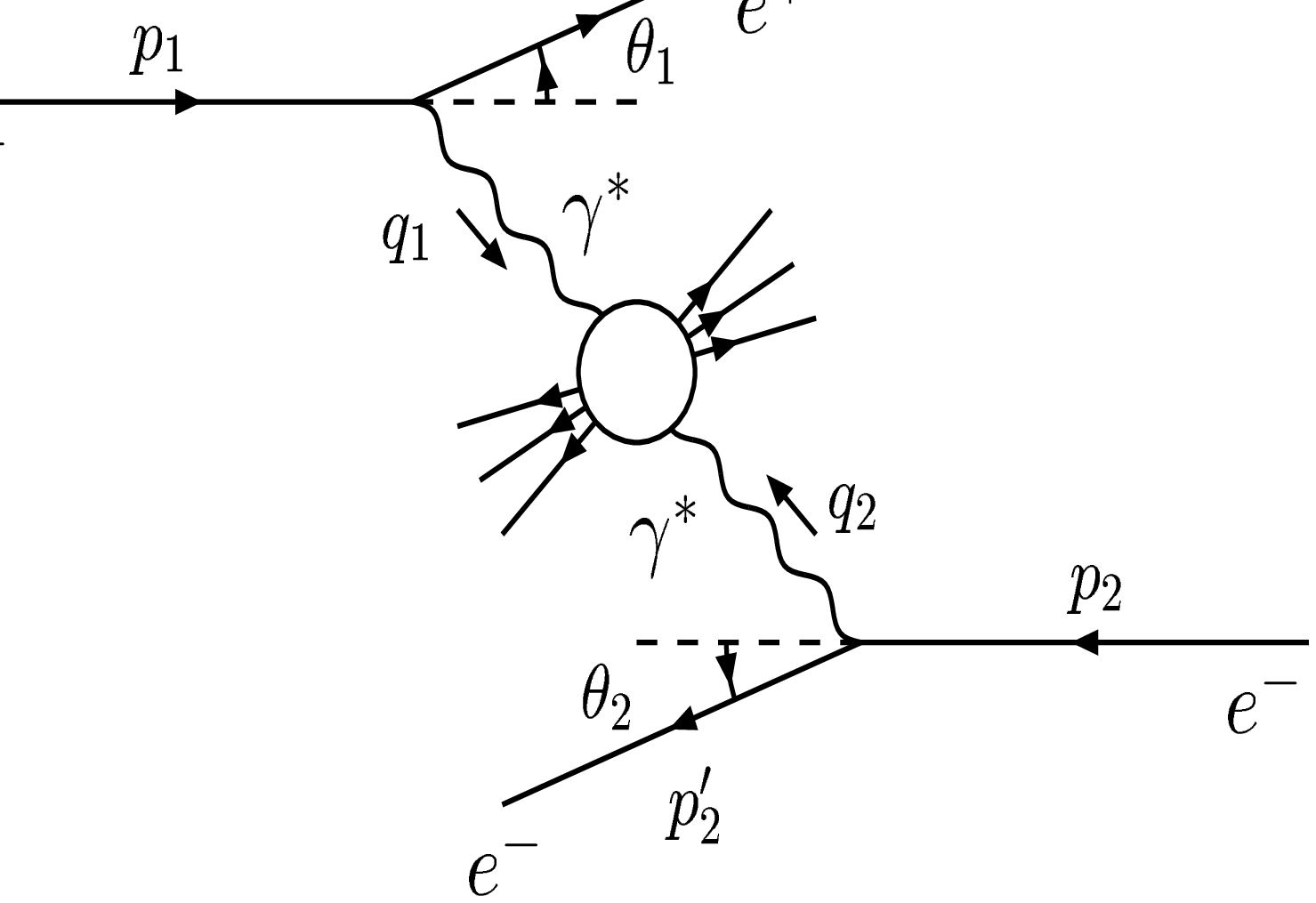,height=9cm,width=10cm}
\epsfig{file=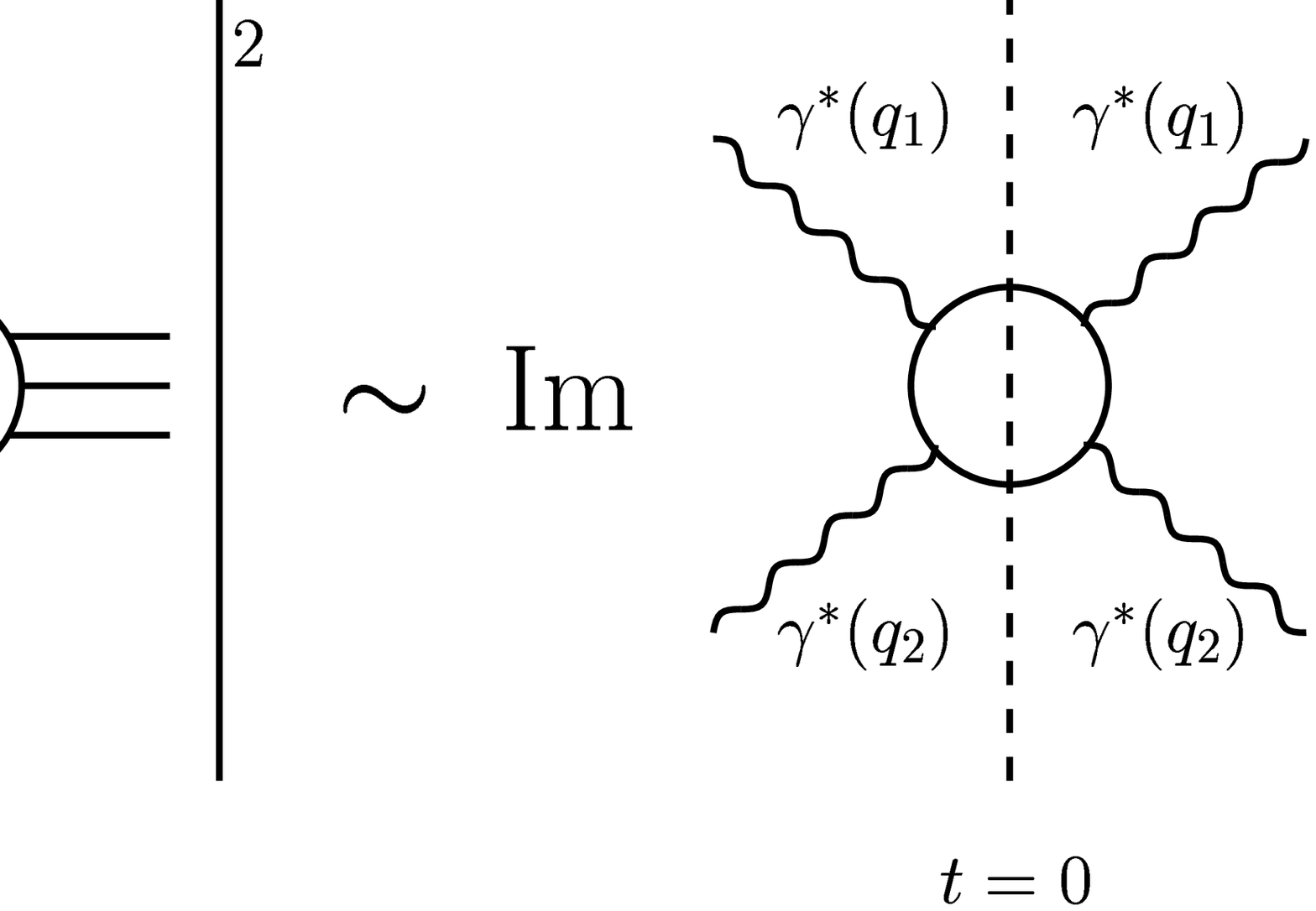,height=9cm,width=8cm}}
\vspace{-4.5cm} 
\caption{  The  $\gamma^*\gamma^*\rightarrow  X$ process 
in $e^+e^-$ collision (left).
 The cross section for the $\gamma^*\gamma^*\rightarrow X$ scattering and its relation to the imaginary part of the 
  forward $\gamma^*\gamma^*\rightarrow \gamma^*\gamma^*$ amplitude (right).  } 
\label{fig1} 
\end{figure} 

One can relate these quantities to the imaginary part of the  forward  
helicity amplitudes $W_{\lambda_1^{\prime} \lambda_2^{\prime}, 
\lambda_1 \lambda_2}$ for the process $\gamma^*(\lambda_1) \gamma^*(\lambda_2)
 \ra 
\gamma^*(\lambda_1') \gamma^*(\lambda_2')$ \cite{bud}, see Fig.3 (right).
Amplitudes with $\lambda_{1(2)}=\lambda'_{1(2)}$  are related to  
corresponding cross sections; the interference terms
correspond to helicity-flip forward amplitudes:
$ \tau_{TT} =  A  W_{++,--}$ and  $\tau_{TL}  = A (W_{++,00} - W_{0+,-0} ) /2$, where  $2\;A = ({(q_1 q_2)^2 - q_1^2 q_2^2})^{1/2} $.

To measure structure functions of virtual photon $\gamma^{\ast}$
the  double-tag events  with $\mid q_1^2 \mid \gg \mid q_2^2 \mid \neq 0$
are used.
Below I will concentrate on  DIS$_{e \gamma^*}$ events with $l^+l^-$
 final state, where
$Q^2\gg P^2\gg\mu^2$ (using a standard notation for  a probe
$q\equiv q_1$ and $Q^2=-q^2$, and for a target $p\equiv q_2$ and $P^2=-p^2$), 
with  $\mu$ - a characteristic scale for  studied phenomena (here $m_l$). 
The   structure functions 
for a polarization-averaged virtual photon target,
 $F_1^{\gamma^*},F_2^{\gamma^*}$ and $F_L^{\gamma^*}=
F_2^{\gamma^*}-2xF_1^{\gamma^*}$ 
 can be introduced, \eg  
\bea
F_2^{\gamma^*}={{Q^2}\over{4\pi^2\alpha}}{{A}\over{2}}
(\sigma_{TT}
+\sigma_{LT}
-{{1}\over{2}}\sigma_{LL}-{{1}\over{2}}\sigma_{TL}).
\eea 
If, after integration of the differential cross section (3)
 over $\bar {\phi}$, the contributions of
the terms  $\tau_{TT}$ and $\tau_{TL}$ vanish, 
one can relate the  measured cross section to  
an effective structure function for a virtual 
photon, $F_{eff}^{\gamma^*} \sim \sigma_{eff}$ \cite{bud}.
 This is {\sl not} the case in the recent OPAL 
measurement of the $\mu$-pair production \cite{OPAL-qed,nisius},
as can be seen in Fig.4. Here  the Monte Carlo predictions based on the  
QED calculations,  
also for the options with  $\tau_{TT}$=0 and $\tau_{TT}$ =
$\tau_{TL}$=0, to test relevance of the interference terms,
are displayed. \\
\begin{figure}[b] 
\centerline{\epsfig{file=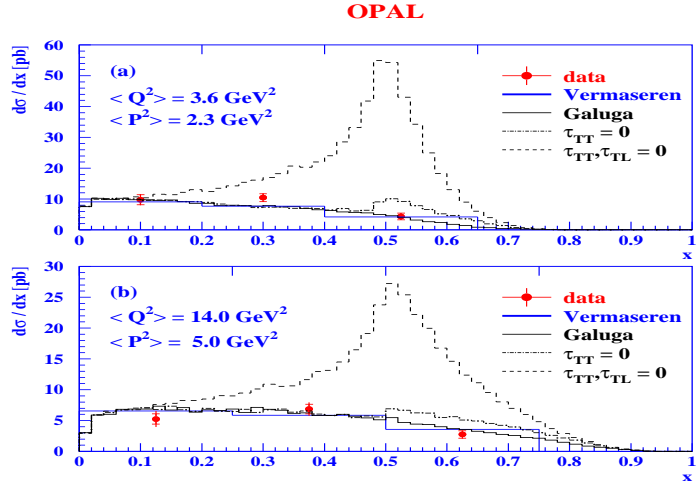,height=7cm,width=7cm}}
\caption{ Differential cross section for $\mu^+\mu^-$
production in $\gamma^*(Q^2)\gamma^*(P^2)$ collision at LEP1 (OPAL Coll.).
The thick line is a QED prediction of
Vermaseren Monte Carlo;  predictions represented by  the  solid line and
dot-dashed (dashed) line were obtained by using  GALUGA MC program for
all contributions and 
assuming $\tau_{TT}=0$ ( $\tau_{TT}$ = $\tau_{TL}=0$), from [29].
}
\end{figure}
Large (negative) intereference terms found for $x>0.1$ 
in double-tag leptonic events at LEP1,   as apparent from Fig. 4, 
make the extracting of the corresponding leptonic structure 
function for $\gamma^*$  unfeasible, and  also shed a light on 
potential problems in extracting the  hadronic structure function for
 $\gamma^*$.
 
Note that for  $P^2\ra 0$  only one interference term, $\tau_{TT}$, 
 remains in cross section (3).
It is called also the  $F_3^{\gamma}$
 structure function \cite{walsh}b).  
In the lowest order QED it has the scaling property,
 $F_3^{\gamma}|_l=- \alpha/\pi Q_l^4 x_{Bj}^3$, so 
 it should not influence  significantly  the extraction of 
 $F_2^{\gamma}$ for a real photon at large $Q^2$.
\subsection*{Hadronic structure function of the  photon}
\subsubsection*{the target = $\gamma$}
In principle one can introduce also ``partonic structure of the photon'' 
since the photon could  materialize itself as a pair of quarks. 
The splitting  $\gamma\rightarrow q\bar q$ leads to the corresponding 
 photon 
structure functions already in the lowest order QED,  
 equivalent here  to the Parton Model (PM).

The Parton Model prediction for the deep inelastic scattering
$e \gamma\ra e + ~hadrons$ is based on the process
 $\gamma^* \gamma \ra q \bar q$. Prediction for the ({\it hadronic}) 
$F_2^{\gamma}|_q$
 is given by the  formula  as for 
a leptonic final state (eqs. 1 and 2), with some modifications:
 $m_l \ra m_q$, $Q_l \ra Q_q$ and  the  color factor, $N_c=3$, has to appear.
 For a final state with  heavy quarks
this is   the  modification (QPM formulae), for light quarks one usually uses
 the massless approximation, with the  QCD parameter, $\Lambda_{QCD}$,
as an argument in the leading logarithm.
So, we have in the Parton Model (in LLA) the following expressions
 for hadronic $F_2^{\gamma}$ and for the (light) {\it quarks densities}:
\bea
F_{2}^{{\gamma}}=
\sum_{q,\bar{q}}{Q_q^2}x_{Bj}q^{\gamma}(x_{Bj},Q^2),\\
q^{\gamma}(x_{Bj},Q^2)={{\alpha}\over{2\pi}}Q^2_q
N_c[x_{Bj}^2+(1-x_{Bj})^2]
\ln{{Q^2}\over {\Lambda_{QCD}^2}}.
\eea

As previously  for leptons, $F_2^{\gamma}$ 
has been  calculated within the QED - it is proportional to $\alpha$!
Both the  $x_{Bj}$ and the $Q^2$ dependence are obtained, 
both are the same as for 
leptonic final state: large $F_2^{\gamma}$ value at large  $x_{Bj}$, 
a logarithmic rise with $Q^2$  (here called a {\it scaling violation}). 
  
The leading logarithmic QCD corrections introduce  logarithmic ($Q^2$) 
modifications
 of the basic predictions of the Parton Model (5),(6), and include also
 a {\it gluonic content} of the photon. 
These corrections  can be  summmed up by
solving the corresponding inhomogeneous  evolution equations. 
By solving them without any input (boundary condition), 
assuming  only that the {\it particular}  solution of the equation
 has the  $Q^2$ dependence as in  PM (eqs. 5-6) 
one obtains  the so called {\it asymptotic solution} \cite{witten}.
It corresponds to the collinear configuration of successive emissions 
of quarks and gluons (as in Fig. 1 (center)), 
all of them   based on the point-like couplings of QED
(the first and the last one), and of QCD.

However, the asymptotic, purly perturbative  solutions suffer from  {\it power singularities}.
For example  the moments ($f^n(Q^2)=\int dx x^{n-1}f(x,Q^2)$)
 of  the non-singlet  structure function are given by  
$f^n(Q^2)|_{asym} \sim \log Q^2/(1- d_n)$, where $d_n$ are proportional to the
$n$th -moment of the PM term: $ [x^2+(1-x)^2]$. 
Simple poles  occur for $d_n=1$, leading after inverting 
the moments to the following  small $x$ behavior of  $F_2^{\gamma}$:
 $(1/x)^{n=1.596}$  (LLA) and $(1/x)^{n=2}$  (NLLA) \cite{bb,uematsu,bar,rossi}.    
The singularities become 
increasingly severe   with higher order of QCD  calculation \cite{rossi}.
Already in the NLLA they lead to the negative value of $F_2^{\gamma}$ \cite{bb}b).

To  cure the problem
of singularities in the (asymptotic) structure functions for a real photon, 
one should include in the calculation also 
the hadron-like ({\it non-perturbative (NP)}) contribution \cite{bar,rossi}. 
The hadronic properties of the photon are apparent  in the {\it soft} 
photon-hadron interaction, where the similarity between photon and 
vector mesons $\rho,\omega,\phi$ interaction is observed 
(VMD model \cite{vdm}).
This NP component  can be included \eg in a boundary
condition at $Q^2_0$ scale, \cite{bar}c,
\be
f^n(Q^2)={{4 \pi}\over{\alpha_s(Q^2)}}[1-({{\alpha_s(Q^2)}\over{\alpha_s(Q^2_0)}})^{1-d_n}]{{\tilde a_n}\over{1-d_n}} 
+ [{{\alpha_s(Q^2)}\over{\alpha_s(Q^2_0)}}]^{-d_n}f_n(Q^2_0),
\ee
where  the input at scale $Q^2_0$ (even $f^n(Q_0^2)$=0!)
 regularizes the  bad behaviour present for 
$d_n\ra 1$, since 
$ {{1}\over{\epsilon}}(1-w^{\epsilon}) \ra -\log w$ for $\epsilon \ra 0$.
By doing this we get rid of the power singularities for the real photon 
structure functions, at the same time we lose an ``absolute'' predictivity 
of QCD for this quantity.

Equation (7) shows why it is customary to treat the structure functions
or the quark densities in the  photon
as being $\sim \alpha/\alpha_s$
although the primary $\log Q^2$ dependence present in the Parton Model
(egs. 5-6), which remains
also after  QCD corrections,  has nothing to do with  $\alpha_s$.
This way of counting changes organization of the perturbation expansion in
 the  QCD calculations, see \cite{order}. 

In some approaches one treats    the 
photon in the hadronic   mode almost as an independent object.
Probing the ``structure'' of the photon in, say, state of $\rho$
can be performed  in an 
analogous way as testing the  structure of other hadrons, 
\eg in the deep inelastic scattering \cite{sas}, see Fig.5.
\begin{figure}[t] 
\centerline{\epsfig{file=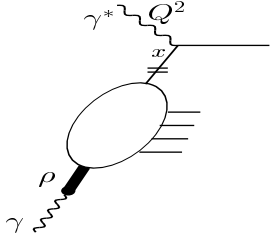,height=6.5cm,width=7cm}} 
\vspace{-2.cm} 
\caption{The deep inelastic scattering on the  photon in the $\rho$
state.}
\end{figure}
The general behaviour of the partonic content of the $\gamma$ in the    $\rho$
mode is known - the  scaling property in the PM and the 
logarithmic scaling violation due to the QCD corrections. 
The corresponding DGLAP 
evolution equation are homogeneous as for the proton, and an input at some
 scale is needed to solve the equation, etc... 
\begin{figure}[t] 
\centerline{\epsfig{file=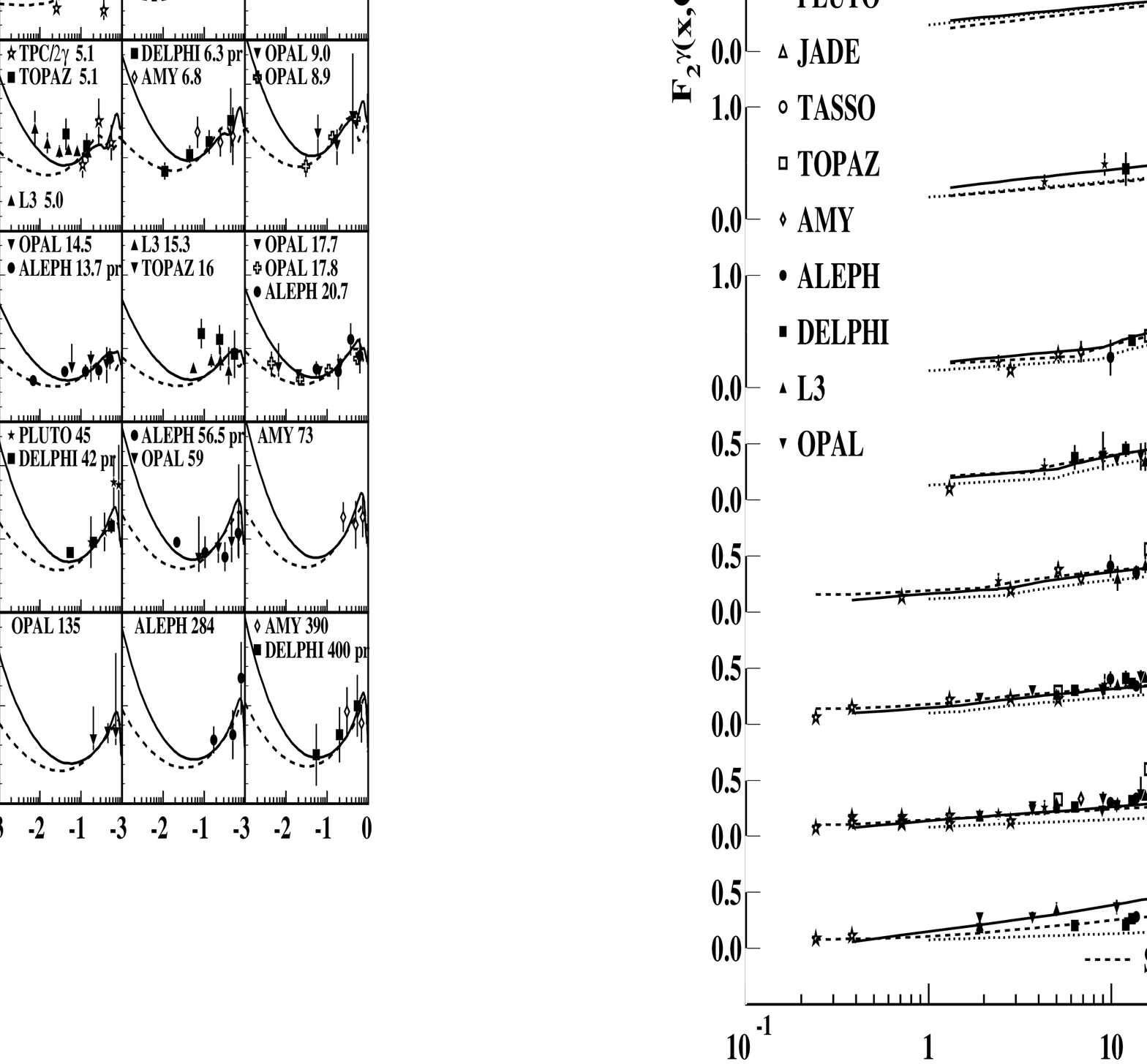,height=9.7cm,width=9cm}}
\vspace{-2.cm} 
\caption{ A compilation of the photon structure function 
$F_2^{\gamma}/\alpha$ data  as a function of $x_{Bj}$ in bins of
$Q^2$ (left) and as a function of $Q^2$ for $<x_{Bj}>$ bins (right)
compared to the GRV NLO (solid line) and 
SaS1D (LO) (dashed line) parametrizations of parton distributions in
the photon, from [33].
} 
\end{figure} 

Various parton parametrizations were constructed for a real photon
in the past (there are about 20 of them, see compilations  in \cite{nisius},
\cite{rev}b).
 The  earlier ones  were based on  a simple Parton Model formula (for quarks) 
or the asymptotic solutions. The 
later parametrizations were 
 based on  approaches  incorporating hadronic-like (NP)  contributions
at some stage. 
Recently parametrizations for a real photon are  obtained 
from   parametrizations   for  virtual photon 
for $P^2\ra 0$\cite{sas},\cite{grv}b).

The compilations of the all existing data for the structure function for
the real photon \cite{stefan} 
are presented in Fig. 5 in comparison with two parton 
parametrizations, obtained in two different approaches to the 
hadron-like contributions,  GRV \cite{grv}a) and SaS \cite{sas}.  
The general behaviour  of the $F_2^{\gamma}$
measured for hadronic  and leptonic  final   states 
is very similar, compare Fig.5 and Fig.2. 
 
An additional  information on the ``structure'' of the photon
is coming from the production of  heavy quarks
 in  photon-induced processes.
  The recent DIS-type measurement at LEP
led to the extraction, for the first time,
of the  charm contribution to $F_2^{\gamma}$,   $F^{\gamma}_{2,c}$, see
\cite{nisius-c}.
The QCD description of heavy quark production
in processes induced by photons is not satisfactory \cite{hq},
 however similar  problem with proper description of 
a heavy quark production exists also in pure hadronic processes.

\subsubsection*{the target = $\gamma^*$}

The structure function of the virtual photon  can 
be calculated  in the Parton Model (QED!) from   
the $\gamma^* \gamma^*\ra q \bar q$ process. 
For $Q^2\gg P^2\gg m_q^2$ ($x_{Bj}=Q^2/2p\cdot q$) one obtains
\begin{eqnarray}
{F_2^{\gamma^*}(x_{Bj},Q^2,P^2)= 
N_c \sum_{q, \bar  q }{{{\alpha}}\over{\pi}}} Q_{i}^4
x_{Bj}[ [x_{Bj}^2+(1-x_{Bj})^2]\ln{{Q^2}\over{P^2x_{Bj}^2}}+
6x_{Bj}(1-x_{Bj})-2],
\end{eqnarray}
to be compared with the eq.~(1). 
 The corresponding PM quark density 
in the virtual photon defined in the LL approximation has the form:
\begin{eqnarray}
q^{\gamma}(x_{Bj},Q^2,P^2) = {\alpha\over 2\pi}  Q_q^2 N_c
[x_{Bj}^2+(1-x_{Bj})^2] \ln {Q^2\over P^2}.
\end{eqnarray}

The QCD  
evolution equations in $Q^2$ for the virtual photon are analogous to those
for the real photon with the inhomogeneous term given by the 
corresponding PM expression.
In the case of the virtual photon one can solve the evolution equation
without the initial conditions. Assuming that 
 for $Q^2\gg P^2\gg \Lambda^2_{QCD}$ the nonperturbative effects
are  absent (see ref.\cite{uematsu}) one obtains for moments of the 
non-singlet  structure function
\be
f^n(Q^2)={{4 \pi}\over{\alpha_s(Q^2)}}[1-({{\alpha_s(Q^2)}\over
{\alpha_s(P^2)}})^{1-d_n}]{{\tilde a_n}\over{1-d_n}}~\sim ~\log
 {{Q^2}\over{P^2}},
\ee
So, without additional experimental or model assumption 
the definite, {\it singularity free}  (asymptotic)
predictions can be derived for both the $x$ and the $Q^2$ dependence 
- a unique situation in QCD. 
Note however that   in all recent analyses  nonperturbative component
in $F^{\gamma^*}_2$ is introduced  \cite{grv}b),\cite{sas}.

Measurements of  the structure functions of $\gamma^*$ can be   
performed in $e^+e^-$ collision, as discussed for a leptonic final state.
The DIS$_{e\gamma^*}$ events with hadronic final state 
were studied experimentally  by PLUTO Coll.\cite{PLUTO},
 new data  
have appeared from  LEP  (L3 Coll.\cite{L3}). In Fig.6
 the  results from both experiments,  
corresponding  to   effective structure functions (cross sections), 
are presented in  comparison with predictions of the QPM, soft VDM, 
GRS \cite{grv}b) models, see however\cite{grv-czy}.

There are already few parton parametrizations for a virtual photon
(see collections in \cite{nisius},\cite{rev}b)), they
are valid for $0\leq P^2$ and become the corresponding
parametrizations for the real photon in the limit $P^2\rightarrow 0$.
All these  parametrizations  deal with the 
{\it {transversely}} polarized {\it virtual} photon, with one 
exception of the Ch\'yla parametrization \cite{chyla2} for a {\it longitudinal}
virtual photon, see also \cite{friburg}.

New insight into the photon structure may come from
spin-dependent structure functions \cite{u}, not measured so far.
Especially the  structure function $g_1^{\gamma}$ is of great importance, 
since its   first moments ({\sl a sum rule for the 
``spin `` of the photon}) involve  strong and electromagnetic anomalies,
and it is deeply connected with the chiral 
properties of QCD \cite{efremov}. It maybe studied at 
future Linear Colliders  \cite{stratman}.

\section*{Conclusion and outlook}
A photon, considered as an ideal probe of hadron structure,
paradoxically   is also considered as an ideal target
to test the perturbative QCD.
Both a real and a virtual photon may reveal ``inner structure'' 
in the interaction with other particles.
 An apparent hadronic structure of the photon 
is clearly seen in the data.
However, there is no full agreement
between the standard QCD predictions 
and experiment for various distributions for hadronic processes
induced both by real and virtual photons. During last years we learned that an improvement of the description of the data can be obtained by introducing  extra $p_t$ for constituents in the target\\
\newpage
\vspace*{13cm}
\begin{figure}[ht]
\vskip -2.1in\relax\noindent\hskip -1.5cm
       \relax{\includegraphics{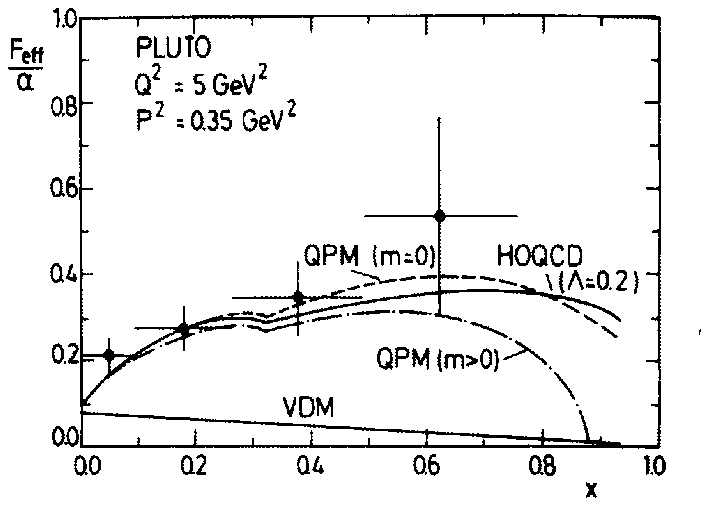}}
\vskip 2.in\relax\noindent\hskip -2.5cm
        \relax{\includegraphics{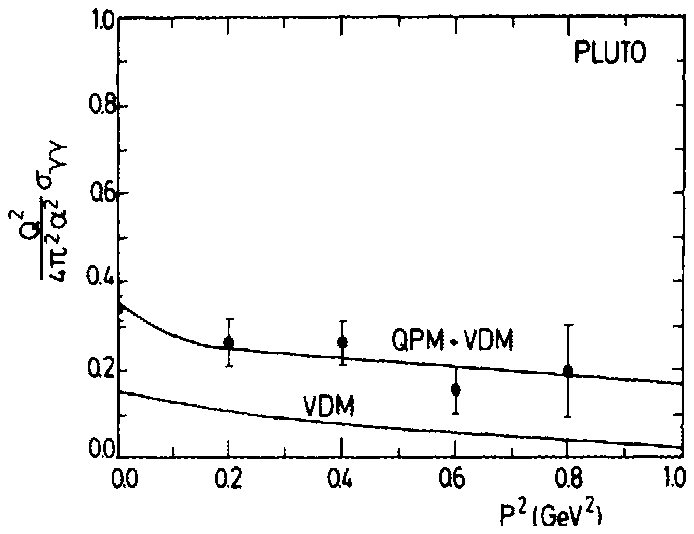}}
\vskip -0.15in\relax\noindent\hskip 6.5cm
       \relax{\includegraphics{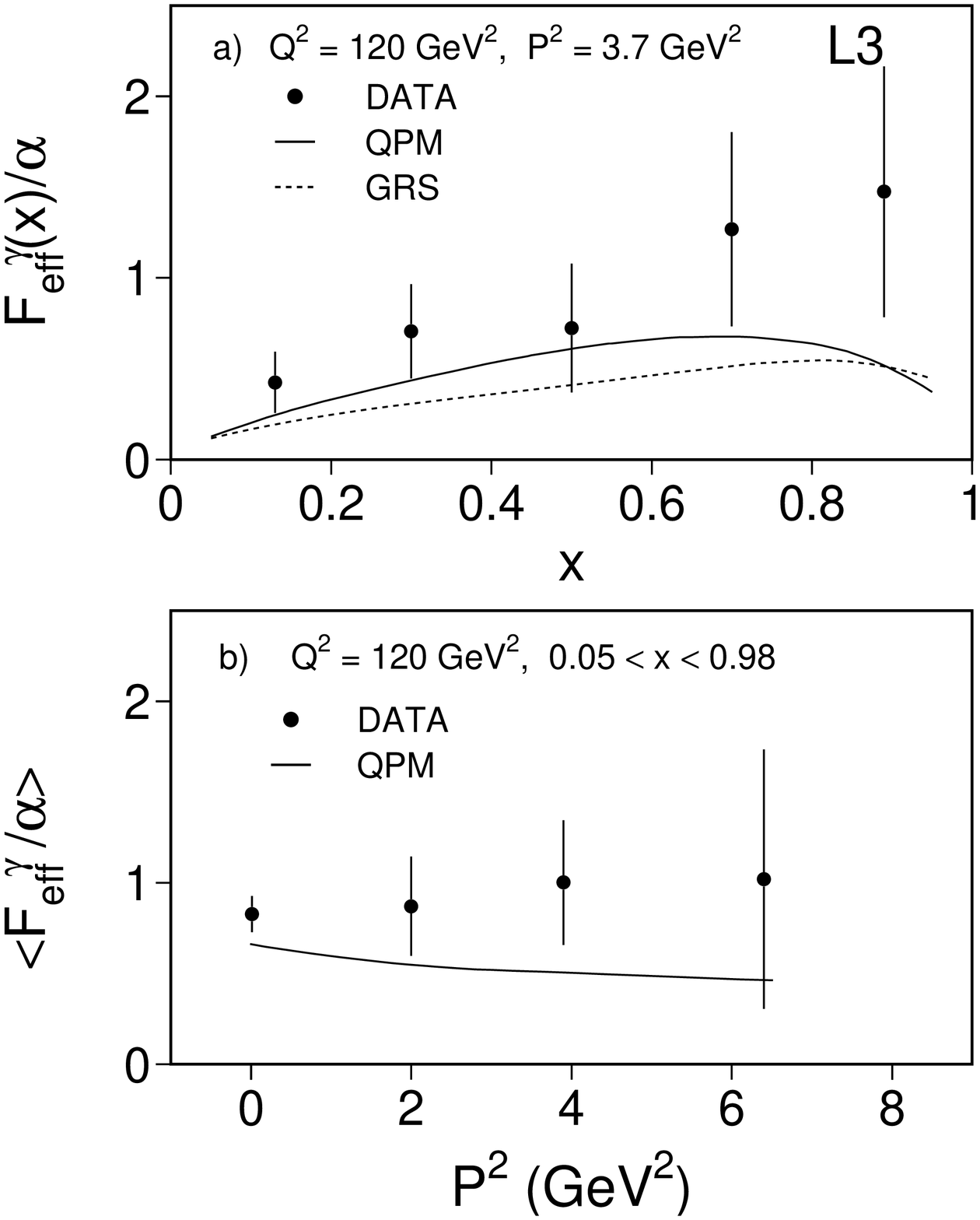}}
\vspace{-3.7cm}
\caption{ The  data for the  $F_{eff}^{\gamma}/\alpha$ 
and comparison with various predictions for the virtual photon.
PLUTO Coll. data for $<Q^2>=5$ GeV$^2$, and $<P^2>=0.35$ GeV$^2$ [27],
and L3 Coll. data  for $<Q^2>$=120 GeV$^2$ and $<P^2>$=3.7 GeV$^2$ [28].
The $x$-dependence (upper panels) and $P^2$-dependence (lower panels)
are presented.}
\end{figure} 
photon, and/or the contributions
 due to  multiple interaction.
Still there are problems in describing
 $x_{\gamma}$ distribution from the dijet events,  prompt photon  
events  in the photoproduction  at HERA, heavy quark production in 
the $e^+e^-$ and $ep$ processes.

For a virtual proton  even more basic questions are open.
How important are   interference terms, 
both for  $\gamma^* \gamma^*$ and $\gamma^* p$ processes? 
Do we see more than just the PM  content of the virtual photon in present 
data? If yes, do we need a ``structure'' of 
the longitudinal virtual photon? 

We have a lot to improve in our description of the photon-hadron interaction.
\section*{Acknowledgments}
I would like to thank organizers of this interesting conference
for fruitful atmosphere.
I am grateful to F.Kapusta, R. Nisus, T. Sj\"ostrand, J. Ch\'yla,
S. S\"oldner-Rembold, D. Miller,  I. Ginzburg,
U. Jezuita-D{a}browska, P. Jankowski and Magda Staszel 
for many illuminating discussions. I appreciate the collaboration with 
U. Jezuita-Dabrowska and P. Jankowski on this contribution.

\end{document}